\documentclass[a4paper,11pt]{article}
\usepackage{pos}

\title{Search for secluded dark matter with 6 years of IceCube data}

\author{The IceCube Collaboration \\{\normalsize \normalfont(a complete list of authors can be found at the end of the proceedings)}}

\emailAdd{ctoennis@icecube.wisc.edu}

\abstract{The IceCube neutrino observatory--installed in the Antarctic ice--is the largest neutrino telescope to date. It consists of 5,160 photomultiplier-tubes spread among 86 vertical strings making a total detector volume of more than a cubic kilometer. IceCube detects neutrinos via Cherenkov light emitted by charged relativistic particles produced when a neutrino interacts in or near the detector. The detector is particularly sensitive to high-energy neutrinos of due to its size and photosensor spacing. In this analysis we search for dark matter that annihilates into a metastable mediator that subsequently decays into Standard Model particles. These models yield an enhanced high-energy neutrino flux from dark matter annihilation inside the Sun compared to models without a mediator. Neutrino signals that are produced directly inside the Sun are strongly attenuated at higher energies due to interactions with the solar plasma. In the models considered here, the mediator can escape the Sun before producing any neutrinos, thereby avoiding attenuation. We present the results of an analysis of six years of IceCube data looking for dark matter in the Sun. We consider mediator lifetimes between 1 ms to 10 s and dark matter masses between 200 GeV and 75 TeV.

\vspace{4mm}
{\bfseries Corresponding authors:}
Christoph T\"onnis$^{1,2*}$

{$^{1}$ \itshape Department of Physics, Sungkyunkwan University, Suwon 16419, Korea}\\
{$^{2}$ \itshape Institute of Basic Science, Sungkyunkwan University, Suwon 16419, Korea}

$^*$ Presenter

\FullConference{37$^{\rm{th}}$ International Cosmic Ray Conference (ICRC 2021)\\
		July 12th -- 23rd, 2021\\
		Online -- Berlin, Germany}

}

\begin{document}
\maketitle

\section{Introduction}

The IceCube Neutrino Observatory~\cite{detector} is to date the largest neutrino telescope. Located at the geographic South Pole, the observatory consists of one cubic kilometer--scale Cherenkov radiation detector built in ice. IceCube has an in-ice cubic kilometre of instrumented volume at depths between 1.450~km and 2.450~km, as well as a square kilometer large cosmic--ray air--shower detector called IceTop~\cite{IceTop} at the surface of the ice. The primary scientific goals of the detector are to measure high-energy astrophysical neutrinos and to identify their sources. Using the data of the IceCube detector a range of dark matter (DM) searches is also being conducted~\cite{Solar,GC,decay}.\\
Although the case for the existence of DM is strong its exact nature remains unknown. There is a variety of candidate models that have been proposed~\cite{Bertone} including weakly interacting massive particles (WIMPs) where the DM particle interacts with Standard Model particles on the scale weak interaction. For indirect searches this type of model is interesting as it gives rise to a flux of Standard Model particles as the result of decays or annihilations of WIMPs. It also leads to the accumulation of DM in massive objects like the Sun or the earth by WIMPs loosing momentum in scattering inside the object and then being gravitationally trapped. \\
The WIMPs accumulated in the Sun decay with the number of accumulated WIMPs $N$ following the Boltzmann equation

\begin{equation}
    \frac{dN}{dt} = C_C - C_A N^2,
\end{equation}

\noindent with the capture rate $C_C$ and the annihilation factor $C_A$. Given this an equilibrium between capture and annihilation will establish itself over timescales of~\cite{Gaisser} 

\begin{equation}
    \tau= \frac{1}{\sqrt{C_C C_A}}.
\end{equation}

In case of such an equilibrium between capture and annihilation the equation

\begin{equation}
    N^2 C_A = 2\Gamma = C_C,
\end{equation}

\noindent holds with the annihilation rate $\Gamma$. Given the age of the Sun with $4.7 \times 10^9$ years the equilibrium is given for this analysis. In this analysis we are searching for neutrinos generated by a type of DM called secluded dark matter that is accumulated in the Sun.\\


\subsection{Secluded dark matter}

\begin{figure}
	\centering
	\includegraphics[width=0.7\textwidth]{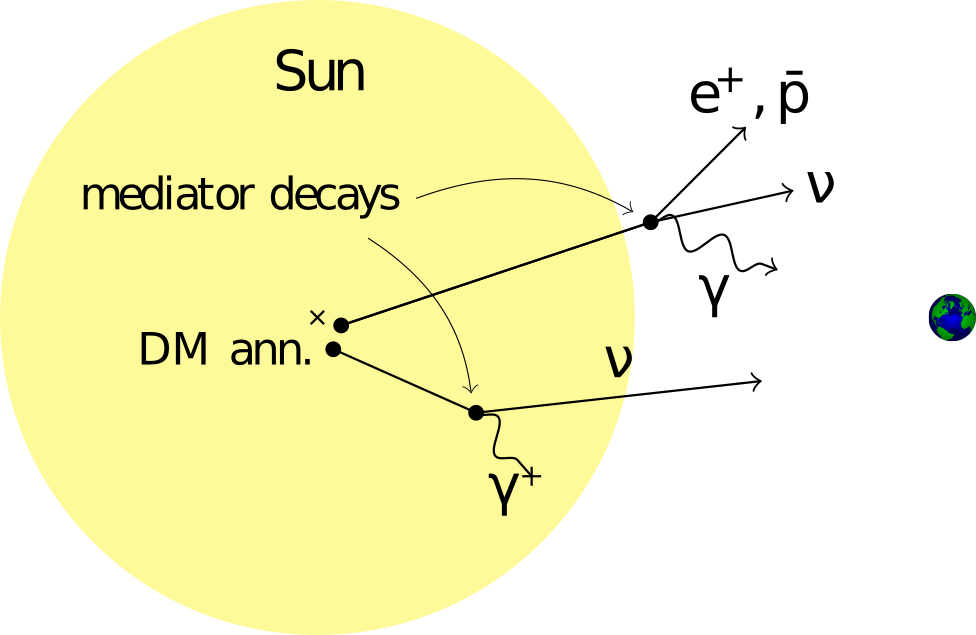}
	\caption{A diagram of secluded dark matter annihilations in the Sun. Two mediators paths are shown: one with a decay length larger than the Sun radius and another mediator decaying inside the Sun.}\label{fig:diagram}
\end{figure}

Secluded dark matter (SDM) is a unique type of model for particle DM where the DM particles do not directly decay or annihilate into Standard Model particles, but rather produce a pair of metastable mediator particles in annihilation that decay after lifetimes that can exceed several seconds into a pair of Standard Model particles. A schematic of this is shown in figure~\ref{fig:diagram}. SDM models arise out of a variety of scenarios for supersymmetric dark matter~\cite{SDM_SUSY} and models of dark photons~\cite{SDM_Photon} or a dark higgs particles~\cite{SDM_Higgs}.  The mediator is not a Standard Model particle and is interacting with Standard Model particles significantly more rarely than neutrinos with ordinary matter.

\begin{figure}
	\centering
	\includegraphics[width=0.7\textwidth]{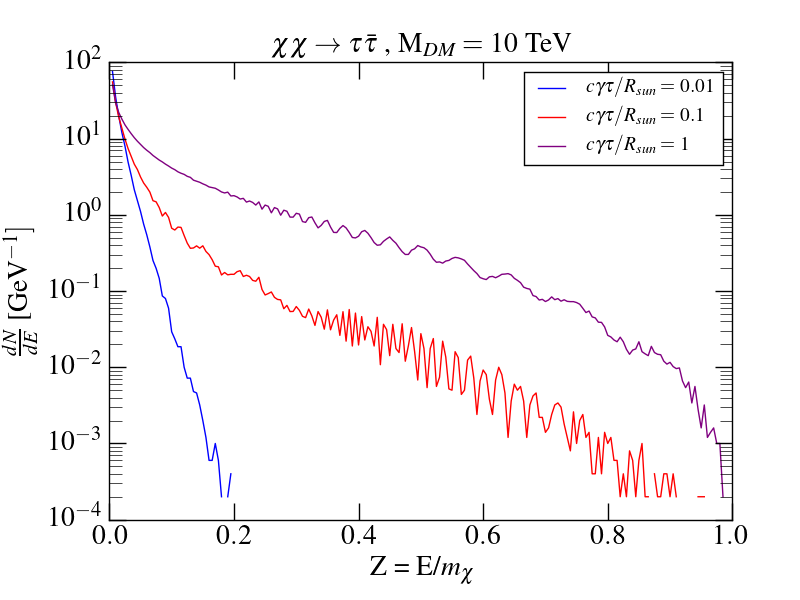}
	\caption{Spectra generated with the WIMPSIM simulation in comparison to the results from the WIMPSIM publication on the 5.0 update of the code~\cite{wimpsim,SWDM}.}\label{fig:spectra}
\end{figure}

SDM is of particular interest for indirect searches towards the Sun. Regular DM models yield signal fluxes from the Sun that are heavily attenuation by the dense solar plasma. This means the neutrinos flux is effectively cut off at energies above 1\,TeV of neutrino energy. However, for SDM this attenuation is avoided when the decay length of the mediator exceeds the radius of the Sun. In these cases the neutrino signal is generated in mediator decays happening outside of the solar plasma and there is no opportunity for neutrinos to interact with the solar plasma. This can be seen in figure~\ref{fig:spectra}, where at longer mediator decay lengths the cutoff at a 10 \% of the dark matter mass, e.g. 1\,TeV, is not present for longer mediator decay lengths.

\section{Analysis method}

For this analysis a sample of muon neutrinos arriving at the detector from below the horizon, e.g. with a zenith angle of more than 90 degrees, recorded from 2011 to 2016 with 1057.8 days of livetime was used~\cite{Reimann}. As neutrinos from above the detector are excluded the Earth is used as a shield against atmospheric muons, which would otherwise comprise a significant background for the analysis. \\
\indent In this search a wide range of model parameters was considered. DM masses ranging from 250~GeV to 75~TeV and mediator decay lengths of 0.01 solar radii to 1 solar radii were considered. Mediator masses ranging from 10 GeV to 10 TeV have been studied. To simulate the expected neutrino signal the WIMPSIM simulation package~~\cite{wimpsim,SWDM} was used. This package does not include some electroweak corrections that are known to have a strong effect on signals from mediators decaying into quarks or neutrinos directly~\cite{Bauer,Charon,Baratella}. Consequently only mediator decays into tau leptons and W bosons were considered in this analysis.\\ 
\indent The sample was analysed using an unbinned likelihood-based method. The likelihood function used in this analysis is defined as

\begin{equation}
    \mathcal{L}(n_s)= \prod_{i=1}^{N_{tot}}\left(\frac{n_s}{N_{tot}} S(\psi_i,E_i) + \frac{N_{tot} - n_s}{N_{tot}} B(\psi_i,E_i) \right),
\end{equation}

\noindent where $n_s$ is the supposed number of signal events, $N_{tot}$ is the total number of events in the sample and $S$ and $B$ are probability density functions (pdfs) describing  the likelihood of the event with an angular separation to the Sun $\psi_i$ and reconstructed energy $E_i$ for the event number $i$. \\
To generate the signal pdf {\it S} the expected neutrino signal was generated using the WIMPSIM package~~\cite{wimpsim,SWDM}. WIMPSIM generates mediators at an annihilation point close to the centre of the Sun that assuming that DM follows a thermal distribution. The mediator is then moved a distance that is based on the assumed decay length of the mediator before simulating its decay.  Using Pythia-6.4.26 the mediator decay into Standard Model particles and the decay and/or hadronisation of these particles is simulated. The resulting primary and secondary neutrinos are then propagated through the remaining solar plasma and the rest of the distance to the Earth including charged current and neutral current interactions with the plasma and vacuum neutrino oscillations.\\
The background pdf {\it B} is estimated using real data by scrambling the right ascension direction of the events. The likelihood is then optimized with respect to $n_s$ and a test statistic value (TS) is calculated as

\begin{equation}
    TS= -2\log\left(\frac{\mathcal{L}(n_{opt})}{\mathcal{L}(0)}\right),
\end{equation}

\noindent where $n_{opt}$ is the value of $n_s$ where the likelihood is maximal. To study the sensitivity of the analysis pseudo experiments (PEs) with varying amounts of inserted signal events were generated and from these TS distributions were calculated for each DM mass, mediator decay channel, mediator decay length and mediator mass and for different values of inserted simulated signal events staring with pure background. The TS distributions are calculated as binned histograms of PEs.\\
Confidence intervals were determined using the Feldman--Cousins method~\cite{fcousins}. For each bin of each TS distribution a rank is calculated. This rank is the ratio of the likelihood to obtain the TS value corresponding to the bin given the actual mean number of inserted simulated signal events used in the distribution, e.g. the bin value, and the likelihood of obtaining the same TS given the best-fit physically allowed number of inserted simulated signal events. Intervals are calculated by adding bins ordered by rank starting with the highest ranked bin until the confidence level a confidence level (CL) of 90\% is reached. Limits were set for the signal strength at which the 90\% CL intervals were entirely at values larger than the TS value calculated for the actual data.
The initial limits are set on average numbers of detected events. These are converted to neutrino flux limits using the detector acceptance $Acc$, which is defined by:

\begin{equation}
    \Phi_{90\%}= \frac{\mu_{90\%}}{Acc},
\end{equation}

with the 90\% CL limit to the total neutrino flux $\Phi_{90\%}$ and the 90\% CL limit in term of average detected signal neutrino events $\mu_{90\%}$. \\
The calculation of the signal acceptance is done using the standard IceCube detector simulation. Simulated events in the data are weighted with the product of the Monte Carlo weight of the simulation and the expected signal neutrino spectra. The weight are computed from a number of parameters such as interaction cross section of neutrinos in the detector volume for interactions that yield detectable neutrino events, the density of the detector volume and the efficiency with which neutrino events are registered in the detector. The same neutrino spectra used to calculate the function $S$ and the IceCube detector simulation is used for the acceptances as well.\\
With the number of neutrinos per annihilation $N_\nu$ calculated from the spectra from~\cite{wimpsim} for each mediator lifetime and DM mass the annihilation rate can be calculated as

\begin{equation}
    \Gamma= \frac{4\pi \rm AU^2 \Phi_{90\%}}{N_\nu},
\end{equation}

where AU is the astronomical unit. With the DarkSusy code package~\cite{conversion,darksusy} the capture rate can the be related to the spin dependent scattering cross section $\sigma_{SD}$. DarkSusy Calculates the capture rate by performing a numerical integration over the solar radius, the velocity distribution and the momentum exchange and considers scattering processes between DM and up to 289 isotopes in the Sun using the Solar model by Serenelli et.al~\cite{Serenelli}. This way the capture rate is expressed as a function of the spin dependent WIMP-nucleon scattering cross section. \\

\section{Results}

\begin{figure}
	\centering
	\includegraphics[width=0.7\textwidth]{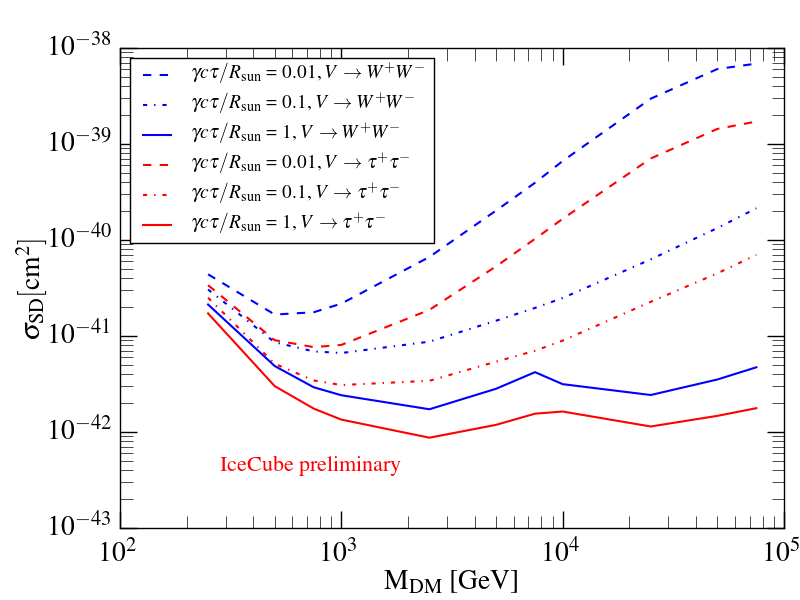}
	\caption{The limits on the spin dependent dark matter-nucleon scattering for different mediator lifetimes In this plot a mediator mass of 100 GeV and a mediator decay length of one solar radius was assumed.}\label{fig:lifetime}
\end{figure}

The analysis has not found any significant indication of SDM in the Sun. Consequently limits on spin dependent scattering cross sections were set. These can be seen for the case of a 100 GeV mediator in figure~\ref{fig:lifetime}. The various DM mass cases are strongly correlated and the likelihood thus finds in most cases no signal events.\\
A kink in the limits can be seen a 7.5 to 10 TeV DM mass at the longest mediator decay lengths. in these cases the signal distribution in reconstructed energy becomes particularly similar to the background causing the likelihood method to fit a small and insignificant number of less than 0.6 signal events. The TS at these masses is still within the 90\% C.L. background intervals, however the limit that can be set here is weaker causing the kink.

\begin{figure}
	\centering
	\includegraphics[width=0.7\textwidth]{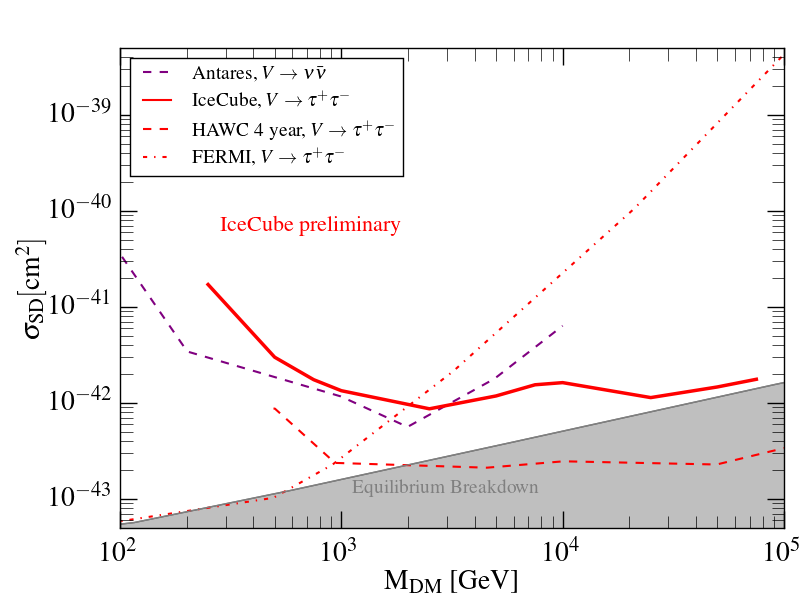}
	\caption{The limits on the spin dependent dark matter-nucleon scattering in comparison to other experiments~\cite{HAWC}\cite{FERMI}\cite{ANTARES}. The results for the $\tau^+ \tau^-$ channel is shown in comparison to other experiments. In this plot a mediator mass of 100 GeV and a mediator decay length of one solar radius was assumed.}\label{fig:limits}
\end{figure}

A comparison to other experiments is shown in figure~\ref{fig:limits}. These results present the thus far strongest limits on SDM from any neutrino experiment. Even the strong limits from the HAWC experiment are being approached. It is to be expected that for mediator decays into neutrinos IceCube would be able to produce the strongest limits. 

For shorter mediator lifetimes it is unlikely that the HAWC experiment would be able to surpass IceCube as the Sun becomes opaque to gamma rays more quickly as to neutrinos. However there are currently no limits from gamma-ray experiments that could be used for such a comparison.

\section{Conclusion}

The results presented here show exclusion limits comparable to those of other neutrino experiment for SDM. If the same decay channels were used results surpassing those of other neutrino experiments could be provided. Although other experiments can present stronger limits on some cases this type of analysis can be expected to yield leading results for other scenarios. These can be explored in a future analysis that will also include more data and use a more accurate signal simulation that includes electroweak corrections.  

\bibliographystyle{ICRC}

\bibliography{main.bib}

\providecommand{\href}[2]{#2}\begingroup\raggedright\begin{thebibliography}{10}

\bibitem{detector}
{\bfseries IceCube} Collaboration, M.~G. Aartsen {\em et~al.}
  \href{http://dx.doi.org/10.1088/1748-0221/12/03/P03012}{{\em JINST}
  {\bfseries 12} no.~03, (2017) P03012}.

\bibitem{IceTop}
{\bfseries IceCube} Collaboration, R.~Abbasi {\em et~al.}
  \href{http://dx.doi.org/10.1016/j.nima.2012.10.067}{{\em Nucl. Instrum. Meth.
  A} {\bfseries 700} (2013) 188--220}.

\bibitem{Solar}
{\bfseries IceCube} Collaboration, M.~G.~A. et~al.
  \href{http://dx.doi.org/10.1140/epjc/s10052-017-4689-9}{{\em EPJC} {\bfseries
  77} (2017) 146}.

\bibitem{GC}
{\bfseries IceCube} Collaboration, M.~G.~A. et~al.
  \href{http://dx.doi.org/10.1140/epjc/s10052-017-5213-y}{{\em EPJC} {\bfseries
  77} (2017) 627}.

\bibitem{decay}
{\bfseries IceCube} Collaboration, M.~G.~A. et~al.
  \href{http://dx.doi.org/10.1140/epjc/s10052-018-6273-3}{{\em EPJC} {\bfseries
  78} (2018) 831}.

\bibitem{Bertone}
G.~Bertone, D.~Hooper, and J.~Silk
  \href{http://dx.doi.org/10.1016/j.physrep.2004.08.031}{{\em Phys. Rept.}
  {\bfseries 405} (2005) 279--390}.

\bibitem{Gaisser}
T.~K. Gaisser, G.~Steigman, and S.~Tilav
  \href{http://dx.doi.org/10.1103/PhysRevD.34.2206}{{\em Phys. Rev. D}
  {\bfseries 34} (1986) 2206}.

\bibitem{SDM_SUSY}
S.~P. Martin {\em Adv. Ser. Direct. High Energy Phys.} {\bfseries 18} (1998) 1.

\bibitem{SDM_Photon}
B.~Holdom {\em Physics Letters B} {\bfseries 166} (1986) 196--198.

\bibitem{SDM_Higgs}
{B. Batell, M. Pospelov and A. Ritz, Y. Shang} {\em Phys. Rev. D} {\bfseries
  81} (2010) 075004.

\bibitem{wimpsim}
{C. Niblaeus, A. Beniwal and J.Edsj\"o} {\em JCAP} {\bfseries 11} (2019) 011.

\bibitem{SWDM}
A.~R. Maxim~Pospelov and M.~B. Voloshin {\em Phys.Lett.B} {\bfseries 662}
  (2007) 53--61.

\bibitem{Reimann}
{\bfseries IceCube} Collaboration, M.~G.~A. et~al. {\em Eur. Phys. J. C}
  {\bfseries 79} (2019) 234.

\bibitem{Bauer}
C.~W. Bauer, N.~L. Rodd, and B.~R. Webber.

\bibitem{Charon}
{Q. Liu, J. Lazar, C. A. Argüelles and A. Kheirandish} {\em JCAP} {\bfseries
  10} (2020) 043.

\bibitem{Baratella}
P.~Baratella, M.~Cirelli, A.~Hektor, J.~Pata, M.~Piibeleht, and A.~Strumia
  \href{http://dx.doi.org/10.1088/1475-7516/2014/03/053}{{\em JCAP} {\bfseries
  1403} (2014) 053}.

\bibitem{fcousins}
{G. J. Feldman, R. D. Cousins} {\em Phys.Rev.D} {\bfseries 57} (1998)
  3873--3889.

\bibitem{conversion}
{C. Rott, T. Tanaka and Y. Itow} {\em JCAP} {\bfseries 09} (2011) 029.

\bibitem{darksusy}
{T. Bringmann, J. Edsj\"o, P. Gondolo, P. Ullio and L. Bergstr\"om} {\em JCAP}
  {\bfseries 033} (2018) 1807.

\bibitem{Serenelli}
{A. Serenelli, S. Basu, J. W. Ferguson and M. Asplund} {\em Astrophys. J.}
  {\bfseries 705} (2009) L123–L127.

\bibitem{HAWC}
{\bfseries HAWC} Collaboration, A.~A. et~al. {\em Phys. Rev. D} {\bfseries 98}
  (2018) 123012.

\bibitem{FERMI}
{\bfseries Fermi LAT} Collaboration, M.~A. et~al. {\em Phys. Rev. D} {\bfseries
  84} (2007) 03.

\bibitem{ANTARES}
{\bfseries ANTARES} Collaboration, S.~A.-M. et~al. {\em JCAP} {\bfseries 05}
  (2016) 016.

\end{thebibliography}\endgroup

\clearpage
\section*{Full Author List: IceCube Collaboration}




\scriptsize
\noindent
R. Abbasi$^{17}$,
M. Ackermann$^{59}$,
J. Adams$^{18}$,
J. A. Aguilar$^{12}$,
M. Ahlers$^{22}$,
M. Ahrens$^{50}$,
C. Alispach$^{28}$,
A. A. Alves Jr.$^{31}$,
N. M. Amin$^{42}$,
R. An$^{14}$
% ...
\\

\noindent
% ...
$^{12}$ Universit{\'e} Libre de Bruxelles, Science Faculty CP230, B-1050 Brussels, Belgium \\
% ...
$^{14}$ Department of Physics and Laboratory for Particle Physics and Cosmology, Harvard University, Cambridge, MA 02138, USA \\
% ...
$^{17}$ Department of Physics, Loyola University Chicago, Chicago, IL 60660, USA \\
$^{18}$ Dept. of Physics and Astronomy, University of Canterbury, Private Bag 4800, Christchurch, New Zealand \\
% ...
$^{22}$ Niels Bohr Institute, University of Copenhagen, DK-2100 Copenhagen, Denmark \\
% ...
$^{28}$ D{\'e}partement de physique nucl{\'e}aire et corpusculaire, Universit{\'e} de Gen{\`e}ve, CH-1211 Gen{\`e}ve, Switzerland \\
% ...
$^{31}$ Karlsruhe Institute of Technology, Institute for Astroparticle Physics, D-76021 Karlsruhe, Germany  \\
% ...
$^{42}$ Bartol Research Institute and Dept. of Physics and Astronomy, University of Delaware, Newark, DE 19716, USA \\
% ...
$^{50}$ Oskar Klein Centre and Dept. of Physics, Stockholm University, SE-10691 Stockholm, Sweden \\
% ...
$^{59}$ DESY, D-15738 Zeuthen, Germany \\
% ...

\subsection*{Acknowledgements}

\noindent
USA {\textendash} U.S. National Science Foundation-Office of Polar Programs,
U.S. National Science Foundation-Physics Division,
U.S. National Science Foundation-EPSCoR,
Wisconsin Alumni Research Foundation,
Center for High Throughput Computing (CHTC) at the University of Wisconsin{\textendash}Madison,
Open Science Grid (OSG),
Extreme Science and Engineering Discovery Environment (XSEDE),
Frontera computing project at the Texas Advanced Computing Center,
U.S. Department of Energy-National Energy Research Scientific Computing Center,
Particle astrophysics research computing center at the University of Maryland,
Institute for Cyber-Enabled Research at Michigan State University,
and Astroparticle physics computational facility at Marquette University;
Belgium {\textendash} Funds for Scientific Research (FRS-FNRS and FWO),
FWO Odysseus and Big Science programmes,
and Belgian Federal Science Policy Office (Belspo);
Germany {\textendash} Bundesministerium f{\"u}r Bildung und Forschung (BMBF),
Deutsche Forschungsgemeinschaft (DFG),
Helmholtz Alliance for Astroparticle Physics (HAP),
Initiative and Networking Fund of the Helmholtz Association,
Deutsches Elektronen Synchrotron (DESY),
and High Performance Computing cluster of the RWTH Aachen;
Sweden {\textendash} Swedish Research Council,
Swedish Polar Research Secretariat,
Swedish National Infrastructure for Computing (SNIC),
and Knut and Alice Wallenberg Foundation;
Australia {\textendash} Australian Research Council;
Canada {\textendash} Natural Sciences and Engineering Research Council of Canada,
Calcul Qu{\'e}bec, Compute Ontario, Canada Foundation for Innovation, WestGrid, and Compute Canada;
Denmark {\textendash} Villum Fonden and Carlsberg Foundation;
New Zealand {\textendash} Marsden Fund;
Japan {\textendash} Japan Society for Promotion of Science (JSPS)
and Institute for Global Prominent Research (IGPR) of Chiba University;
Korea {\textendash} National Research Foundation of Korea (NRF);
Switzerland {\textendash} Swiss National Science Foundation (SNSF);
United Kingdom {\textendash} Department of Physics, University of Oxford.

\end{document}